\pdfoutput=1
\documentclass{sigchi-ext}
\usepackage[T1]{fontenc}
\usepackage{textcomp}
\usepackage[scaled=.92]{helvet} 
\usepackage{graphicx} 
\usepackage{balance}  
\usepackage{booktabs} 
\usepackage{ccicons}  
\usepackage{ragged2e} 

\def\plaintitle{NaMemo: Enhancing Lecturers' Interpersonal Competence of Remembering Students' Names} 
\def\emptyauthor{}
\def\plainkeywords{Name-indicating system; classroom; university students; teachers' interpersonal competence; user acceptance; HCI.}

\title{NaMemo: Enhancing Lecturers' Interpersonal Competence of Remembering Students' Names}

\numberofauthors{5}
\author{%
  \alignauthor{%
    \textbf{Guang Jiang}\\
    \textbf{Mengzhen Shi}\\
    \affaddr{School of Telecommunication Engineering} \\
    \affaddr{Xidian University, Xi'an, China} \\
    \email{gjiang@mail.xidian.edu.cn} 
	\email{mzshi@stu.xidian.edu.cn} }
	\alignauthor{%
	\textbf{Ying Su}\\
	\affaddr{Suicide Research and Prevention Center}\\
	\affaddr{Shanghai Mental Health Center}\\
	\affaddr{School of Medicine}\\
	\affaddr{Shanghai Jiao Tong University}\\
	\email{suy@smhc.org.cn} } \vfil
 	\alignauthor{%
    \textbf{Pengcheng An}\\
    \affaddr{Department of Industrial Design}\\
    \affaddr{Eindhoven University of Technology}\\
    \affaddr{the Netherlands}\\
    \email{p.an@tue.nl} }
	\alignauthor{%
	\textbf{Yunlong Wang}\\
	\textbf{Brian Y. Lim}\\
	\affaddr{School of Computing}\\
	\affaddr{National University of Singapore}\\
	\email{yunlong.wang@nus.edu.sg}\\
	\email{brianlim@comp.nus.edu.sg} } 
}

\definecolor{linkColor}{RGB}{6,125,233}
\hypersetup{%
  pdftitle={\plaintitle},
  pdfauthor={\emptyauthor},
  pdfkeywords={\plainkeywords},
  bookmarksnumbered,
  pdfstartview={FitH},
  colorlinks,
  citecolor=black,
  filecolor=black,
  linkcolor=black,
  urlcolor=linkColor,
  breaklinks=true,
}

\begin{document}


\maketitle

\RaggedRight{} 

\begin{abstract}
  
  Addressing students by their names helps a teacher to start building rapport with students and thus facilitates their classroom participation. However, this basic yet effective skill has become rather challenging for university lecturers, who have to handle large-sized (sometimes exceeding 100) groups in their daily teaching. To enhance lecturers' competence in delivering interpersonal interaction, we developed NaMemo, a real-time name-indicating system based on a dedicated face-recognition pipeline. This paper presents the system design, the pilot feasibility test, and our plan for the following study, which aims to evaluate NaMemo's impacts on learning and teaching, as well as to probe design implications including privacy considerations.
\end{abstract}

\keywords{\plainkeywords}

\ccsdesc[500]{Human-centered computing~Human computer interaction (HCI)}
\ccsdesc[100]{Human-centered computing~User studies}

\printccsdesc

\section{Introduction}
Teacher-student interpersonal relationship determines students' perceived closeness \cite{Opdenakker2012} to the teacher and thus influences their engagement and performance in classroom learning \cite{Coll2002,Pianta2009,Pakarinen2010}. A classic yet effective way for teachers to build up closeness to students is to address each individual by their names \cite{Glenz2014,Cooper2017}. However, this skill becomes rather challenging for nowadays university lecturers, given the large number (sometimes exceeding 100) of students in a class. Oftentimes, a lecturer could only remember a few students' names throughout a course period: affording these students better chances for interactive participation in the lecture and subconsciously missing personal engagement with the rest of the students \cite{Wubbels2005}.

To tackle this substantial challenge, we explore a novel technology that augments university teachers' interpersonal competence \cite{Wubbels2006} in teaching large student groups. Namely, we design NaMemo, an AR name-indicating system that displays students' names to facilitate a teacher to address any student by their names during a lecture and eventually improves the class quality and experience of both teachers and students.



\section{Enhancing teachers' Interpersonal Competence}
The framework of Classroom Assessment Scoring System (CLASS) \cite{Hafen2015,Alansari2017} argues for emotional support, classroom organization, and instructional support as three key aspects of classroom quality. Regarding emotional support, a sufficient teacher-student rapport \cite{Wubbels2005} is vital to effective and enjoyable learning processes. Addressing students by their names is the classic yet effective starting point of establishing such a rapport \cite{Glenz2014,Cooper2017}.

Research shows that students can benefit from perceiving that instructors know their names \cite{Cooper2017}. Knowing a student's name conveys the feeling that the teacher cares the student and helps develop a sense of trust \cite{Yang2012}; besides, it could increase the chance that the teacher addresses the student by name \cite{Wu2012,Yang2012}. However, it becomes especially challenging in university lecture settings, where nowadays teachers have to cope with rather large classes (sometimes over 100 students in one class). 

This challenge hinders a teacher from building a sufficient rapport with the students and affording each of them adequate opportunities for active participation in the lecture. We are therefore motivated to extend this interpersonal competence of teachers through an interactive system.

\section{Prior Designs for Supporting Teacher-Student Interaction in Classrooms}

Much progress has been made in the design of tools to 
enhance learners' individualized \cite{Deslauriers2019,VanLehn2011,Verbert2013}, gamified \cite{Dicheva2015}, or immersive \cite{Lui2014} learning experiences.

Relatively, a smaller-but burgeoning-body of work has explored designs to enhance teacher-learner interpersonal interaction in the classroom. Examples encompass orchestration tools \cite{Dillenbourg2010}, or peripheral data displays to ease teachers' interpersonal support \cite{Holstein2018}, communication \cite{Verweij2017}, or proximity \cite{An2019} to students. For instance, ClassBeacons \cite{An2019} uses ambient lamps in the classroom to depict how the teacher divides time and attention over pupils in class to enhance teachers' real-time reflection on this important interpersonal competence. Such systems are expected to 
enable effective learning environments by promoting teachers'
multi-modal \cite{Martinez2019} and interpersonal interactions with students. In this paper, we aim to contribute relevant insights  in this burgeoning area through the design and implementation of NaMemo.

\section{The Design of NaMemo System}

\begin{marginfigure}[0.5pc]
	\begin{minipage}{\marginparwidth}
		\centering
		\includegraphics[width=1\marginparwidth]{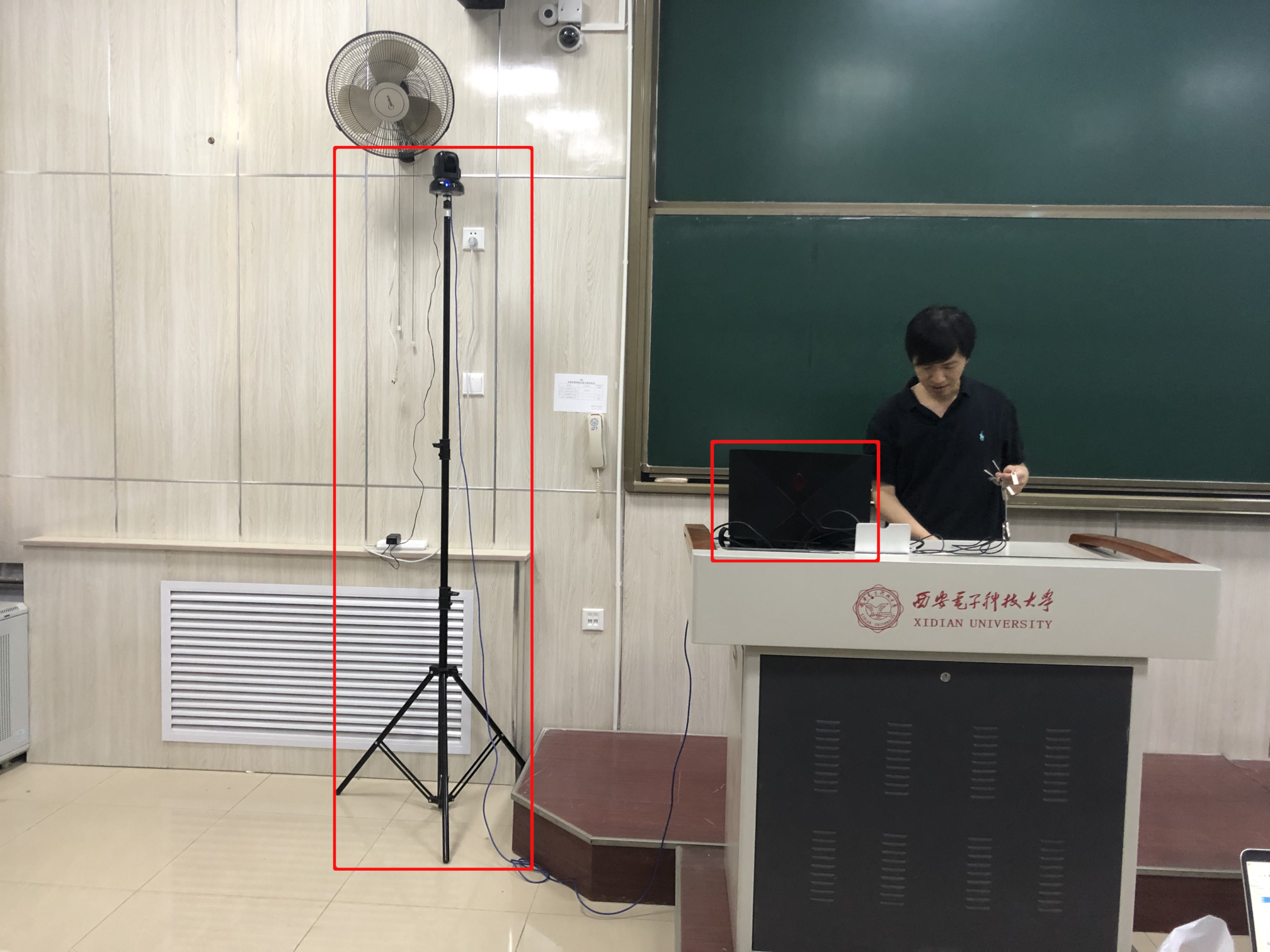}
		\caption{The prototype of our NaMemo system in a real classroom setting, highlighted in red frames.}\label{fig:ClassroomSetting}
	\end{minipage}
\end{marginfigure}

\begin{marginfigure}[0pc]
	\begin{minipage}{\marginparwidth}
		\centering
		\includegraphics[width=1\marginparwidth]{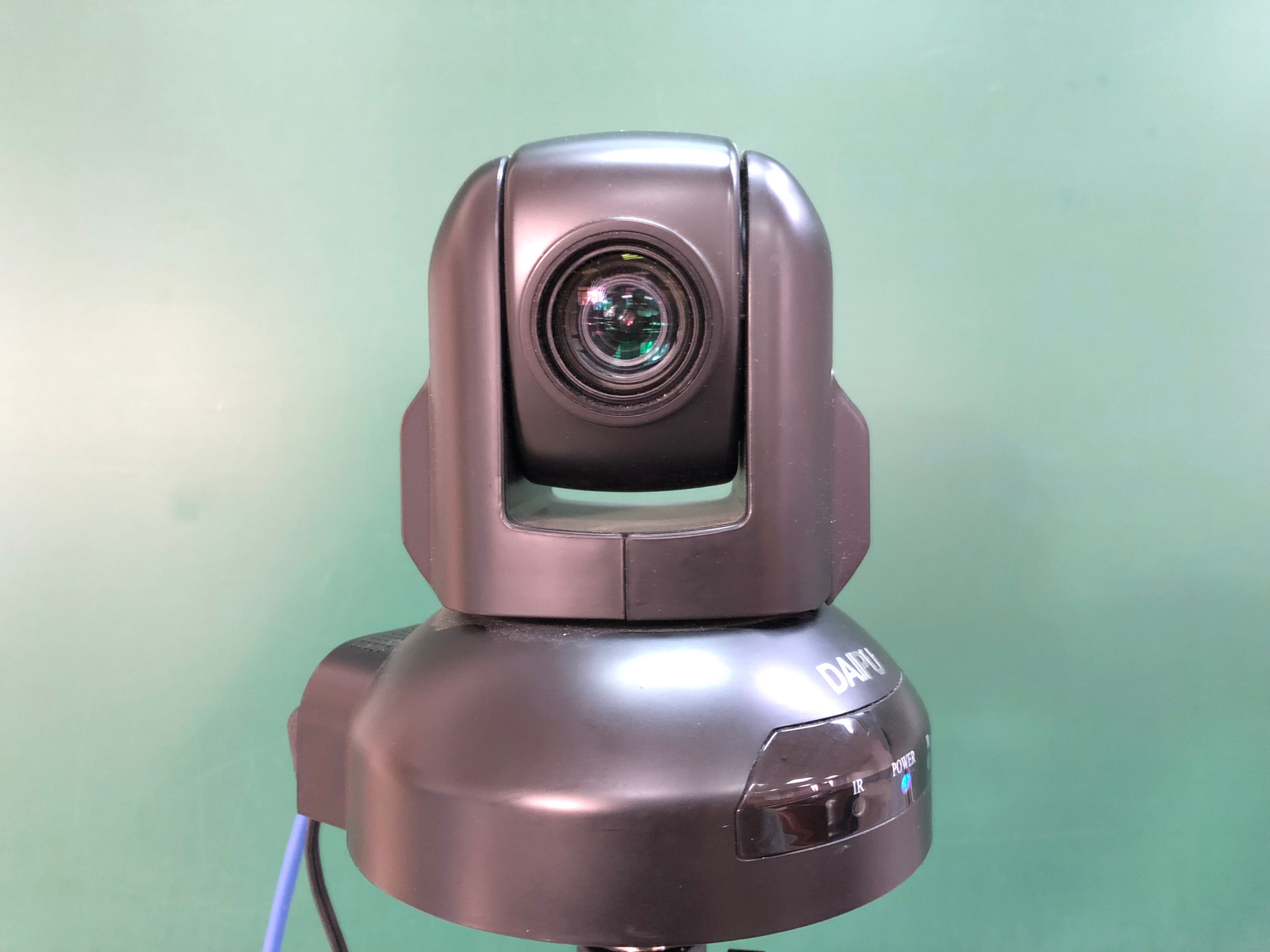}
		\caption{The pan-tilt camera used in our NaMemo system.}\label{fig:Cam}
	\end{minipage}
\end{marginfigure}

The system consists of a pan-tilt camera (about 50 dollars) fixed on a tripod and a laptop (see Figure \ref{fig:ClassroomSetting} and Figure \ref{fig:Cam}). During the class, the system automatically recognizes students and indicates the corresponding names on the screen of the teacher's laptop.

\begin{figure}
	\includegraphics[width=.9\columnwidth]{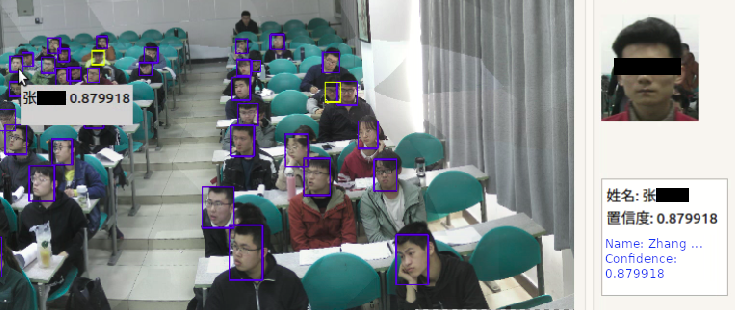}
	\caption{The main user interface of the NaMemo System. The rectangles frame the recognized students. The blue ones indicate confidence levels above 0.8, while that in the yellow ones are below 0.8 and over 0.5. When the cursor is located in a frame area, the teacher could see the student's profile on the right part of the window. The blue text is the translation.}\label{fig:UI}
\end{figure}

To obtain clear images of all areas in the classroom (especially the back rows which are far away from the camera), we choose a camera with a 35mm lens. However, this setting cannot cover the whole classroom with a single image. Therefore, we use a pan-tilt module to rotate the camera and take multiple images (e.g., 63 images in our feasibility test covering a 20m x 15m classroom). Given these images from different rotation angles, we adopt image-stitching and face recognition techniques, resulting in a panorama with student recognition, as shown in Figure \ref{fig:UI}. The panorama is updated every 90 seconds.

We build our recognition model on a convolutional neural network integrating the state of the art techniques for face recognition (i.e., mtCNN \cite{Zhang2016} and ArcFace \cite{Deng2018}). The pipeline in our approach includes three steps: face detection, face alignment, and recognition by comparing with the database. Our approach reaches  99.2\% accuracy on the face recognition task in our feasibility test with dense population and large spatial distribution. The contribution of the algorithm is not the focus of this paper, and we will present the details elsewhere.

\section{Pilot Feasibility Test}
As a pilot, NaMemo was deployed in a university classroom in China. 
A teacher participant used the prototype during his lecture "Micro-controllers: Theory and Applications" with 161 students (in two classes) for ten lectures. All the students were aware of the system's deployment. No video was recorded during the pilot. The pilot study was approved by the local ethics committee in the university. The participants consented and were allowed to opt-out anytime during the pilot. After using the prototype for ten lectures, we used informal (ad-hoc) questionnaires and interviews (with the teacher and ten students) to roughly probe its feasibility and preliminary user experiences. 159 {(99\%)} informal questionnaires were gathered from the students.

The questionnaire results may imply the feasibility of the system:
151 students {(95\%)} were satisfied with the deployment of
the system. As a reference of how the teacher interacted with the
students during the pilot, 97 students were called and all of their names were used correctly. Students' also reported their experiences of the pilot lectures: e.g., perception of the teacher being more approachable {(75\%)}, or themselves being more concentrated during lectures {(78\%)}. Although this might imply some potential benefits of the system, they are to be further verified or falsified in future research.

%
%
%


During the interview, we asked the teacher and ten randomly
selected students about their opinions and concerns about the
system. The teacher thought NaMemo could "improve the in-class
efficiency" by replacing his task of checking a name list to call 
some students' names, which can interrupt his workflow. The Interviewed students felt they became more focused during the class session (knowing that the teacher might call their names), and they felt they could be more active in the class ("I might be able to have more interactions with the teacher"). Although the students were informed that system was only used to indicate names, some students still asked if it was used for checking attendance. After extra explanation, the interviewed students relieved the concern. This suggests the importance of further probing the privacy issues and design solutions for such systems \cite{Ogan2019}.

\section{Discussion and Future Work}



A formal, larger-scale evaluation of NaMemo will be carried out in the future, in which the following aspects could be critically examined:

\textbf{Teacher-Student Interaction.} Our direct goal of developing the NaMemo system is to increase teacher-student interaction in class. In our next study, a between-subject set-up will be adopted to study the effects of NaMemo on two types of interaction: the teacher calling students' name to ask a question (or for other reasons), and students asking questions to (or requesting attention from) the teacher.
  
\textbf{Classroom Learning-Quality.} Ultimately,we are interested in how the system influences students' learning quality, which could be moderated by teach-student interaction. As found by Deslauriers et al. \cite{Deslauriers2019}, students' perceived learning-quality could be different from their actual learning-quality. Therefore, we will measure students' learning-quality both objectively and subjectively \cite{Deslauriers2019}.

\marginpar{%
	\vspace{-85pt} 
	\fbox{%
		\begin{minipage}{1\marginparwidth}
			\section{Acknowledgments}
			\vspace{1pc} 
			G.Jiang and M. Shi developed the system and conducted the study. Y. Wang and P. An led the writing of the paper. All of the authors contributed to the writing and reviewing of the paper. Y. Wang is the corresponding author of the paper.
	\end{minipage}}\label{sec:ack} }

\textbf{Teachers' Memorization of Students' Names.} Besides
indicating students' names on the spot, we wonder whether the system could also improve teachers' memorization of students' names via more frequent interaction with students, or, whether it only offloads teachers' memory and raises their dependency. 

\textbf{Students' Privacy.} Importantly, we plan to critically probe the privacy concerns and ethical issues of using such computer vision systems in classrooms \cite{Ogan2019,Ahuja2019}. While related reflections have concerned the western classroom contexts \cite{Ogan2019,Ahuja2019}, fewer insights are generated in eastern classroom contexts. Since no standard instrument is known for privacy evaluation of classroom technologies, we will develop ad-hoc questionnaires, based on related instruments (e.g., students' perceived control from the teacher from QTI \cite{Wubbels2014}).

\textbf{Education Experts' Opinions.} Besides the perception of the students and teachers, the opinions from education experts are necessary to deepen the insights of our study. We plan to cooperate education experts in our study to evaluate the potential benefits and risks of the system in different perspectives.

\textbf{Using Smart Glass as the User Interface.} In future development, we would like to explore integrating NaMemo system with a smart glasses interface. Using smart glasses, teachers could 
get the digital information simultaneously when observing the classroom, without moving their focus away from the students 
(e.g., see the Lumilo project  \cite{Holstein2019}). 

%
%

\section{Conclusion}
The proposed NaMemo system is in its early stage of exploring technologies supporting teacher-student interaction in class. Therefore, the system is simplistic, and extra functions remain to be implemented. In our future work, we aim to contribute to the HCI community by designing and evaluating new interfaces based on the system to support teachers' interpersonal competence during classroom teaching.




\balance{} 

\bibliographystyle{SIGCHI-Reference-Format}{\small}
\bibliography{my}

\end{document}